# Two-photon pumped exciton-polariton condensation

**NADAV LANDAU,[1,*] DMITRY PANNA,[1] SEBASTIAN BRODBECK,[2] CHRISTIAN SCHNEIDER,[3] SVEN HÖFLING,[2] AND ALEX HAYAT[1]**

*[1]Department of Electrical and Computer Engineering, Technion – Israel Institute of Technology, Haifa, 3200003, Israel*
*[2]Technische Physik, Universität Würzburg, Am Hubland, D-97074 Würzburg, Germany*
*[3]Institute of Physics, Carl von Ossietzky Universität Oldenburg, D-26111 Oldenburg, Germany*
**Corresponding author: nlandau@campus.technion.ac.il*

**Two-photon absorption (TPA) allows accessing "dark" states of matter otherwise inaccessible to light, which serve as important building blocks for quantum information processing. In a semiconductor microcavity, TPA driven condensation of strongly-coupled light-matter exciton-polaritons can enable new solid-state quantum simulations of "dark" state-condensate interactions, and was predicted to stimulate THz emission. Here, we report the first observation of two-photon pumped polariton condensation, demonstrated by angle-resolved photoluminescence in a GaAs-based microcavity. TPA is evidenced in the quadratic emission dependence on pump power below and above the condensation threshold, and second-harmonic generation is ruled out by both this threshold behaviour and by the emission peak energy showing no dependence on pump photon energy. Our results pave the way towards novel polariton-based sources and solid-state coherent control of collective quantum states with individual two-level systems.**

Over the past several decades, progress in high-intensity laser amplifiers and high-quality material fabrication has led to a surge in both fundamental and applied research in the field of nonlinear optics [1]. The discovery of two-photon absorption [2-4] in particular has led to remarkable advances in fields such as optical data storage [5], fluorescence microscopy [6] and 3D nanofabrication [7], while the investigation of quantum-confined structures using multiphoton absorption spectroscopy [8] has revealed a rich variety of so called "dark" material excitations, inaccessible by linear optical methods. Included in this group are the "dark" exciton bound states in semiconductor Quantum Wells (QWs) [9, 10], quantum dots [11, 12] and two-dimensional materials [13], which due to their relatively long lifetimes and reduced dephasing, serve as important building blocks for solid-state quantum information processing [10, 12, 14]. In a strongly-coupled light-matter system of exciton-polaritons, the 2p "dark" exciton state in the cavity-embedded QW can interact with the 1s "bright" exciton-based polaritons via a dipole-allowed 2p-to-1s THz transition [15, 16]. Coherent coupling of this state to the Lower Polariton (LP) and Upper Polariton (UP) resonances has been demonstrated [17], forming a unique three-level system which could allow for novel coherent control schemes in the solid-state [18]. Resonantly driving the 2p exciton state by Two-Photon Absorption (TPA) generates a coherent "dark" exciton population, enabling realizations of additional schemes [4, 10, 19]. Establishing polariton condensation [20-22] at the LP ground state through TPA could then open the door to a new regime of light-matter interaction by coherently coupling optically-controllable collective states of matter to single qubits. This scheme not only creates new possibilities for ultrafast nonlinear optics with polariton condensates, currently relying on slower inter-particle interactions [23], but can also enable new solid-state quantum simulations. It allows for the simulation of state preparation and decoherence mechanisms of collective quantum states in rapidly developing quantum technologies such as superconducting qubits in solid-state systems [24], and Schrodinger cat states in quantum optics [25]. Moreover, it was theoretically predicted that placing such a device inside a THz cavity could lead to a doubly-stimulated emission process – the LP condensate in this case playing the role of an additional bosonic stimulating final-state [15, 26].

Lately, significant efforts have been made towards achieving TPA-based polariton condensation, peaking with the demonstration of two-photon generated non-condensed polaritons [19, 27-31]. However, condensation via TPA was not achieved previously due to the challenging conditions required: on the one hand, the condensation process calls for extremely high intensities injecting large carrier densities by TPA, while on the other, low pulse repetition rates are necessary to allow cool down in-between pulses [32], all while still maintaining strong light-matter coupling in a condensate-supporting structure.

Here we demonstrate for the first time two-photon pumped polariton condensation, achieved by TPA-based excitation at extremely high intensities and low repetition rates. Our pump conditions enable reaching the threshold density via TPA and establishing condensation without breaking the strong light-matter coupling. We show this in a planar GaAs-based microcavity, by pumping with ultrafast pulses at energies near half the exciton levels. The resulting photoluminescence (PL) from the LP ground state exhibits a clear intensity threshold as a function of increased TPA intensity, coinciding with an interaction-induced blueshift and a spectral linewidth narrowing, characteristic of the transition from a polariton thermal distribution to polariton condensation. The emission is found to depend quadratically on pump power below and above the condensation threshold, as expected for a TPA process. The existence of a threshold rules out second-harmonic generation (SHG) as the reason for the quadratic dependence, and we verify this by scanning the pump photon energy and observing a lack of dependence of the emission peak energy.

In our experiments, the output of a femtosecond-pulsed laser was tuned close to half the resonance energy of the 2p exciton state in our sample – at $E_{pump} \cong 800$ meV [Fig. 1]. Optical selection rules around $k_{\parallel}=0$ dictate that in such a TPA process, only "dark" states which are dipole-forbidden from directly coupling to light can be excited [2, 9, 19]. Recent studies have shown that some direct two-photon injection of 1s "bright" exciton-based polaritons is possible in GaAs due to valence-band mixing effects at finite in-plane momenta [29, 30]. However, since the strength of this mixing is very weak for typical photonic wavenumbers, we expect the dominant absorption channel in our excitation scheme to be the 2p "dark" exciton. Macroscopic occupation of the LP ground state and polariton condensation is thus established via several possible relaxation channels, namely the 2p-to-LP THz transition [33] and finite $k_{\parallel}$ phonon-assisted and polariton-polariton relaxation processes [30,32] [Fig. 1]. In our sample, we further exclude excitation of Light-Hole (LH) exciton states as these lie ~35 meV higher than the discussed Heavy-Hole (HH) exciton states, which is greater than both the UP-LP splitting and the spectral region covered by our pump pulse.

PL from our sample was collected from the central area of the two-photon pump spot and far-field analysis was then performed for various 1s exciton-cavity detuning energies Δ. Fig. 2 shows our results for the angle-resolved PL measurements at a slightly negative 1s exciton-cavity detuning energy of $\Delta \cong -4$ meV, and for different TPA powers. Superimposed on the plots are the calculated in-plane dispersions of the type shown in Fig. 1, determined using an appropriate coupled oscillator model [34] fitting our sample parameters.

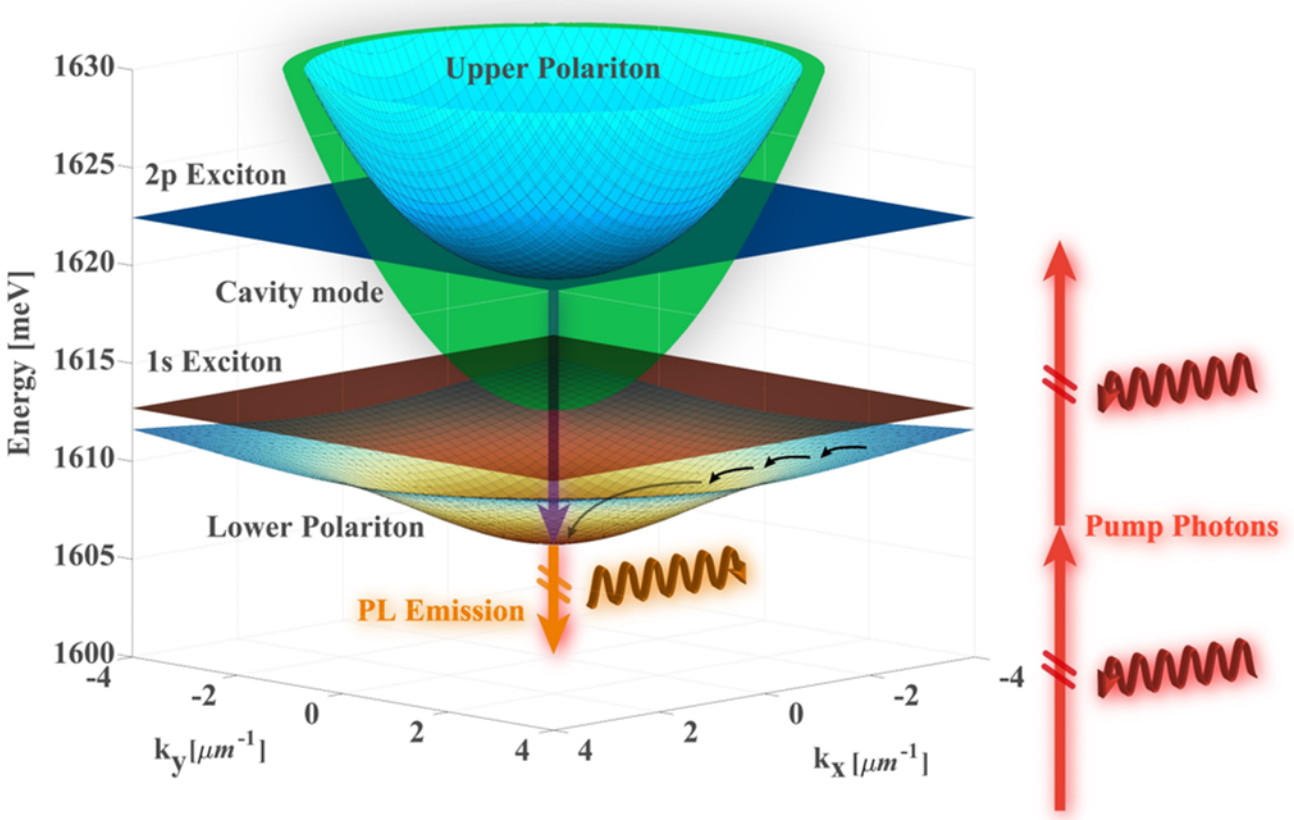

**Fig. 1.** Schematic of the in-plane polariton dispersion relations calculated at a 1s exciton-cavity detuning energy of $\Delta \cong 0$meV, showing the lower and upper polariton branches, as well as the 2p exciton [dark blue], 1s exciton [brown], and cavity photon mode [green] dispersions. The pump photon energy [red] is approximately half that of the exciton states; The LP PL [orange arrow] is also depicted along with the possible THz transition [purple arrow] and phonon-polariton and polariton-polariton relaxation processes on the LP branch [small black arrows]. Photon energies are not drawn to scale, as indicated by the double-diagonal lines on the arrows.

For TPA powers below the condensation threshold, emission from thermally-distributed polaritons on the LP branch is observed, as it is exclusively populated at low temperatures [Fig. 2 (a)]. At an increased power slightly above threshold, a transition appears from the broad-angle thermally-populated LP branch to a narrow and slightly blueshifted state [Fig. 2 (b)]. As has previously been established for a one-photon excitation scheme [35], the above transition is associated with the build-up of a polariton condensate [36,37].

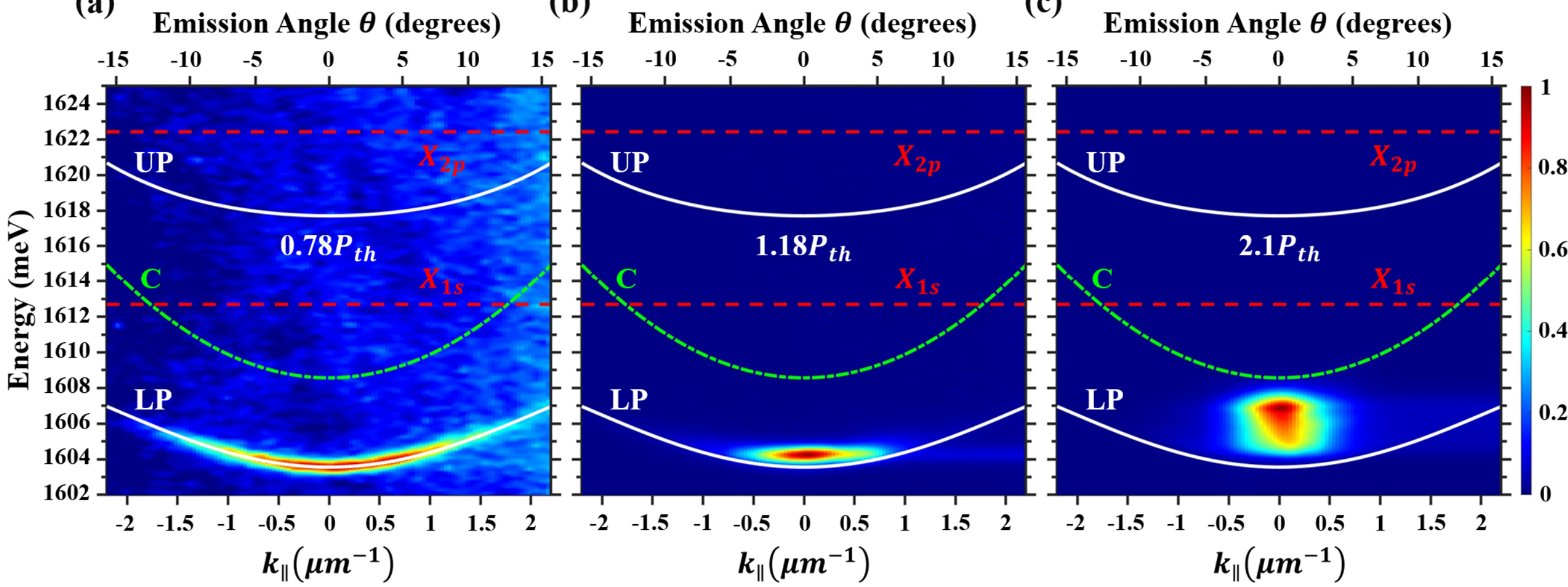

**Fig. 2.** Angle-resolved PL at a 1s exciton-cavity detuning energy of $\Delta \cong -4$ meV, measured at 6K for different TPA powers. The PL in each panel is normalized and plotted in a linear color scale, with the color code indicated on the right. The lines represent the calculated LP and UP (white solid), cavity photon mode C (green dashed-dotted), and 1s ($X_{1s}$) and 2p ($X_{2p}$) exciton (red dashed) in-plane dispersions. **(a)** Below threshold, the emission is broadly distributed in momentum and energy owing to thermally-distributed LPs. **(b)** Slightly above the condensation threshold, the emission narrows spectrally and shrinks to within a small angular range of $\pm\sim3^{\circ}$, originating from slightly blueshifted condensed polaritons around $k_{\parallel}=0$ **(c)** Above threshold, the emission coming from further blueshifted condensed polaritons is smeared due to the time-integrated nature of the measurement; Horizontal axes in each panel display the in-plane momentum $k_{\parallel}$ (bottom) and corresponding emission angle $\theta$ (top).

Increasing the TPA power further above threshold, the emission originates from a now further-blueshifted condensate [Fig. 2 (c)]. This emission is slightly smeared to lower energies because the polariton density changes in time following the ultrashort TPA pulse, so that in the time-integrated image, the blueshift temporal decay appears simultaneously with the peak of the blueshifted state.

To confirm the transition to a polariton condensate from the angle-resolved images, the evolution of several PL characteristics with increasing TPA powers was extracted. Fig. 3 (a) shows the integrated PL intensity at a 1s exciton-cavity detuning energy of Δ ≅-4 meV as a function of the TPA power, where only low in-plane momenta ($|k_{\parallel}|\leq 0.35\mu m^{-1}$) has been considered.

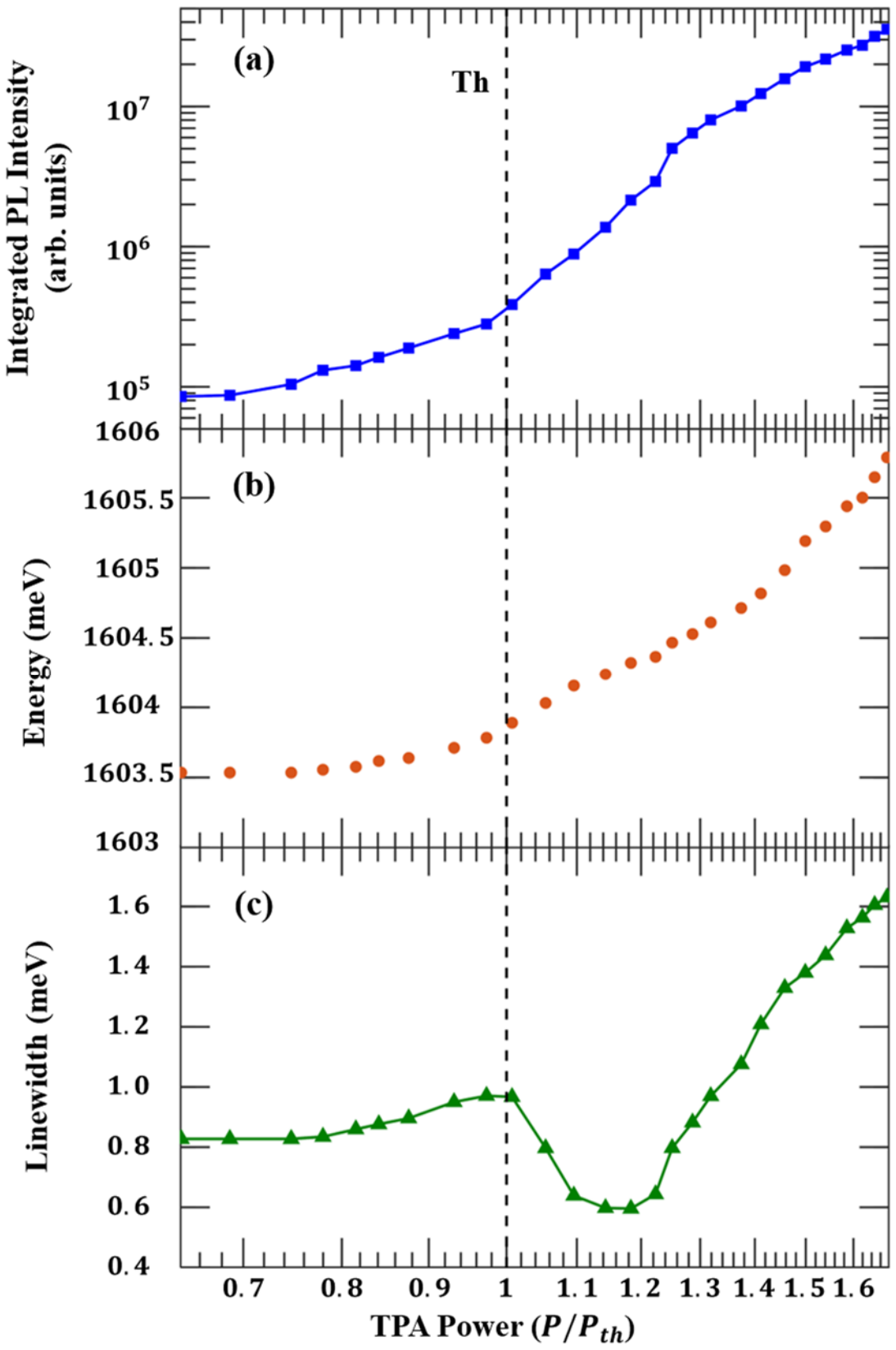

**Fig. 3.** PL characteristics vs. TPA power ($P/P_{\text{th}}$ , logscale) at Δ≅-4 meV. **(a)** the integrated PL intensity (arb. units, logscale) within in-plane momenta $|k_{\parallel}|\leq 0.35\mu m^{-1}$. The quadratic dependence below threshold strongly steepens at the condensation threshold. **(b)** the PL peak energy (in meV) within the same in-plane momenta window. Being roughly constant at the LP ground state energy for lower powers, it continuously blueshifts with increasing power starting from slightly below threshold, due to self-interactions of an increasing polariton density [38,39]. **(c)** the PL linewidth (FWHM, in meV). A clear spectral narrowing at threshold corresponds to increased temporal coherence, and the broadening at larger powers is attributed to density fluctuations and polariton-polariton interactions [40]; The vertical dashed line across all panels indicates the condensation threshold.

Below the condensation threshold, in the thermal regime, the integrated PL intensity increases quadratically with the TPA power [see Fig. 4 (a)], until it abruptly steepens around threshold. This abrupt change occurs at a TPA power similar to the transition observed in the time-integrated image [Fig. 2 (b)].

Tracking the dependence of the PL peak energy on the TPA power, a small blueshift is observed starting from slightly below the marked threshold, as shown in Fig. 3 (b). In the thermal regime, the emission peak energy remains roughly constant at the LP ground state energy of ∼1603.6 meV. Closer to the marked threshold, this peak begins to continuously blueshift with increasing TPA power, owing to the enhanced self-interactions of an increasing polariton density [38,39]. This blueshift is found to be approximately quadratic in pump power for this regime and thus linear in injected population density, as expected for a polariton-polariton interaction-based mechanism following TPA. It further constitutes an important indicator for the presence of a self-interacting polariton condensate, and its total magnitude remains lower than the Rabi splitting.

Examining the evolution of the PL spectral linewidth with increasing TPA powers [Fig. 3 (c)], a very gradual increase of its FWHM bandwidth is initially observed, followed by a sharp narrowing around the condensation threshold down to about two thirds of the value at the thermal regime. This spectral narrowing corresponds to the increased temporal coherence of the condensed state. Increasing the excitation further above the threshold power, the observed linewidth starts increasing continuously due to density fluctuations and polariton-polariton interactions [40], in good agreement with previous observations on one-photon pumped condensation [35].

To further verify condensation in our sample, we have performed the same measurements at two different energy detunings of Δ≅0 meV and Δ≅+4.5 meV between the 1s exciton and cavity mode, and similar characteristic behaviours were observed. All three detunings lie within the previously determined range for establishing polariton condensation in a GaAs-based microcavity [41].

To confirm the two-photon nature of the excitation, Fig. 4 (a) shows the log-scale input-output curves for two of these cases and for an extended TPA power range. Power-law fits below and above

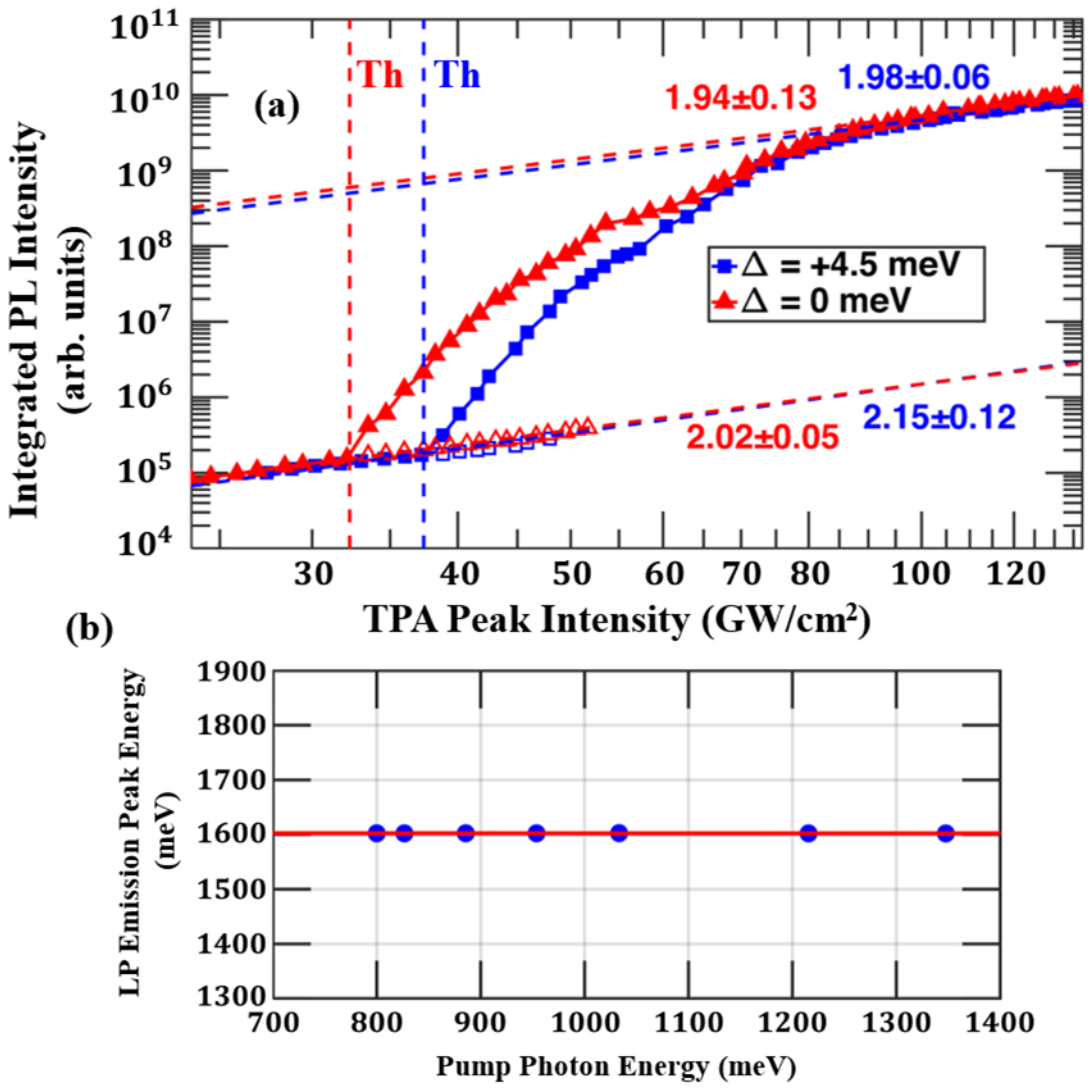

**Fig. 4.** ***(a)*** Integrated PL intensity (logscale, arb. units) vs. TPA peak intensity (logscale, in GW/cm²) for two 1s exciton-cavity detuning energies of Δ≅0meV (red line and triangles) and Δ≅+4.5meV (blue line and squares). Vertical dashed colored lines mark the condensation threshold in each case, of ∼32.3GW/cm² and ∼37.4GW/cm²

respectively. Diagonal dashed colored lines are power-law fits to $y=a\cdot x^b$ where the values of b are 2.02±0.05 and 2.15±0.12 below threshold, for $\Delta\cong 0$meV and $\Delta\cong +4.5$meV respectively, and 1.94±0.13 and 1.98±0.06 above threshold for the same, all in good agreement with a power-of-two input-output dependence, as expected for a TPA process. **(*b*)** LP emission peak energy (in meV) vs. pump photon energy (in meV). Blue circles are recorded data and red line is a linear fit showing that effectively no dependence of the emission peak energy is observed when scanning the pump photon energy across a wide spectral range.

each threshold are also presented. The fitted values for the slopes at each detuning all show good agreement with a power-of-two input-output dependence.

The observed threshold behaviour of the emission rules out direct SHG as the reason for the power-of-two dependence, and to further confirm this we have scanned the pump photon energy and recorded the resulting LP emission peak energy. For a direct pump-induced SHG process, the emission peak energy is expected to shift spectrally along with the scanned pump photon energy. As Fig. 4 (b) clearly reflects no such dependence of the emission peak energy on the pump photon energy, and as mentioned, the emission shows a clear condensation threshold, direct SHG is ruled out as a cause for the observed power-of-two dependence. We can also safely rule out second-harmonic generation followed by one-photon absorption into polaritons, as this is a third-order perturbation process followed by a first-order one – compared to a much stronger second-order perturbation process of direct TPA [1,4].

In conclusion, we observe two-photon pumped condensation of exciton-polaritons. The angle-resolved photoluminescence from the LP ground state in our planar GaAs-based microcavity exhibits a clear, detuning-dependent intensity threshold with increasing TPA peak intensities, at which the onset of a steep nonlinearity in the input-output curve, an interaction-induced blueshift of the central emission peak, and a sharp spectral narrowing of this peak corresponding to an increased temporal coherence, are all simultaneously observed, as expected for a transition to a polariton condensate [41]. The log-scale input-output curves for different 1s exciton-cavity detuning energies exhibit a power-of-two dependence below and above the threshold region, confirming the TPA excitation scheme. The above results present a new platform for the investigation of nonlinearly-excited "dark" states and solid-state single qubits interacting with collective states of matter, with potential applications in solid-state coherent control and quantum information processing. They further generate new possibilities for ultrafast polariton-based nonlinear optics and could advance the realization of a polariton-based stimulated THz radiation source [15].